\documentclass[journal=apchd5,manuscript=draft]{achemso}
\usepackage[T1]{fontenc}
\usepackage[english]{babel}
\usepackage{amsmath,amsthm,amssymb,latexsym,amsfonts,gensymb} %math and symbol packages
\usepackage[printonlyused,nolist]{acronym} %package for abbreviations
\usepackage{graphicx,float,epstopdf,hyperref} %packages (mostly) related to figures

\usepackage{xcolor}
\usepackage[normalem]{ulem}

\title{Tunable photogating in a molecular aggregate coupled graphene phototransistor}

\author{Abhinav~Raina}
\affiliation{Institute for Light and Matter, Universit\"at zu K\"oln, Greinstr. 4--6, 50939 K\"oln, Germany}

\author{Maurizio~Sanfilippo}
\affiliation{Institute for Light and Matter, Universit\"at zu K\"oln, Greinstr. 4--6, 50939 K\"oln, Germany}

\author{Chang-Ki Moon}
\affiliation{Institute for Light and Matter, Universit\"at zu K\"oln, Greinstr. 4--6, 50939 K\"oln, Germany}

\author{Manuel Neubauer}
\affiliation{Institute for Light and Matter, Universit\"at zu K\"oln, Greinstr. 4--6, 50939 K\"oln, Germany}

\author{Klaus~Meerholz}
\affiliation{Institute for Light and Matter, Universit\"at zu K\"oln, Greinstr. 4--6, 50939 K\"oln, Germany}

\author{Malte C. Gather}
\affiliation{Institute for Light and Matter, Universit\"at zu K\"oln, Greinstr. 4--6, 50939 K\"oln, Germany}

\author{Klas~Lindfors}
\affiliation{Institute for Light and Matter, Universit\"at zu K\"oln, Greinstr. 4--6, 50939 K\"oln, Germany}
\email{klas.lindfors@uni-koeln.de}


\begin{document}

\maketitle

\begin{abstract}
We present a graphene photodetector coupled to a layer of aggregated organic semiconductor. A graphene phototransistor is covered with a thin film of merocyanine molecules. The aggregation of the molecular layer can be controlled by the deposition parameters and post-deposition annealing to obtain films ranging from amorphous to a highly aggregated state. The molecular layer has a uniaxial structure with excitonic transitions whose transition dipole moments are well defined. The presence of the molecular layer results in an enormous increase in the response of the phototransistor. We further demonstrate that the signal-enhancement is due to p-photodoping of the graphene. The spectroscopic photoresponse suggests that the photodoping via monomers and molecular aggregates takes place differently. Our photodetector is a platform to study the influence of molecular aggregation and order on charge transport processes between aggregated organic semiconductors and two-dimensional materials.
\end{abstract}

\section*{Introduction}

Graphene attracts significant attention in optoelectronics due to it’s high carrier mobility, zero bandgap, ultra-fast carrier dynamics, and tunable work function.~\cite{Schwierz2010, Bao2012, Avouris2010, Castro_Neto_2009, Ferrari2015} However, the weak light absorption of graphene (2.3\%) makes it not optimal for light harvesting devices such as photodetectors and phototransistors. Despite this, significant efforts have been made to utilize graphene for optoelectronic devices by virtue of conjugating them with various organic or inorganic materials with strong optical responsivity.\cite{Koppens2014,Han2022,Guo2024} Among the inorganic materials, graphene photodetectors have been conjugated with plasmonic nanoparticles to not only significantly enhance the generated photocurrent but also to induce spectral selectivity by virtue of the plasmon resonance of the plasmonic nanoparticles.\cite{Liu2011} Further, PbS quantum dots have also been shown to enhance the photoresponsivity of graphene photodetectors reaching values of 10$^{9}$~AW$^{-1}$.\cite{Konstantatos2012} In this case, the strong light absorption of the inorganic material results in the generation of charge carriers which when transferred to the underlying graphene layer result in a significant enhancement of the photoresponse. While these approaches improve the responsivity of the graphene devices, they offer little in the way of tunability, as the optical responses of such inorganic materials are intrinsically rigid. Here, organic molecules offer an interesting alternative. The chemical nature of graphene makes it an ideally suited template for aromatic organics.\cite{Guo2024,Han2022,Dai2024,Joo2012,Jang2012,Tang2020,Liu2016,Liu2021,Nguyen2015,Lee2011,Han2012,Han2018,Jones2017,Gim2016,Xiong2020} The exposed p$_{z}$ orbitals that lie orthogonal to the graphene layer are ideally suited for strong electronic interactions with such organic molecules. This enables influencing the electronic properties of graphene via the optoelectronic response of the organic component. For instance, studies have used thin-films of photo-switching azobenzene derivatives to reversibly tune the optoelectronic properties of the underlying graphene layer.\cite{Tang2020,Peimyoo2012} Here, the Dirac point of a graphene transistor can be reversibly adjusted by photoswitching of the azobenzene molecule. This shift in the Dirac point is caused by the formation of an interfacial electric field generated by the transfer of photogenerated charge carriers in the organic layer upon optical excitation. Similar effects have also been shown for thin films of the organic dye spiropyran.\cite{Joo2012,Jang2012} Using the graphene layer as a transistor affords the additional opportunity to investigate the dynamics of photogenerated charge carrier transport, both within the organic layer as well as across the interface between the organic layer and graphene, by virtue of the facile tunability of the Fermi level of graphene.

Another interesting prospect is provided by aggregating organic semiconductors. Here, the formation of molecular aggregates causes dramatic shifts not only in the light absorption in thin films of the dyes but also the charge transport properties, which affords additional modalities of tunability to such a hybrid transistor.\cite{Yakar2020,Sun2021} Frequently, molecular aggregation significantly alters the light absorption and charge transport properties of thin films of organic semiconductors. Yakar \emph{et al.}~\cite{Yakar2020} recently demonstrated a J-aggregate/graphene hybrid transistor where a thin film of molecular aggregates was formed on a graphene transistor via membrane casting. The photoexcitation of the thin film can be used to reversibly photodope the graphene layer thereby shifting the Dirac point of the graphene transistor. The authors used a membrane casting technique due to the fact that the aggregating dye (along with most other aggregating organic dyes) was water soluble.  

Here we use the merocyanine dye HB238~\cite{buerckstuemmer2011} in a hybrid organic-graphene phototransistor. This organic semiconductor has been designed for applications in organic photovoltaics and displays very high light absorption. Thin films of HB238 can be deposited by physical vapor deposition (PVD) or spin coating directly on the graphene layer, and its aggregation state can be controlled by the deposition conditions.~\cite{boehner2024} The optoelectronic properties of thin films of the semiconductor are highly dependent on the relative state of aggregation resulting in large spectral shifts as well as significant enhancement in carrier mobility.~\cite{boehner2024,schaefer2024} 
Furthermore, due to the presence of in-and out-of-plane components of the formed oblique angle molecular aggregates, the optical response also varies as a function of angle of incidence. This ease of deposition coupled with the interesting optoelectronic properties of aggregates of HB238 make it an ideally suited candidate for the fabrication of a multimodally tunable hybrid graphene phototransistor. Such a device is a testbed to study the influence of molecular aggregation and order on charge transport at the graphene-organics interface.

\section*{Results and discussion}

\subsection*{Sample fabrication}

\begin{figure}
    \centering
    \includegraphics[width=16cm]{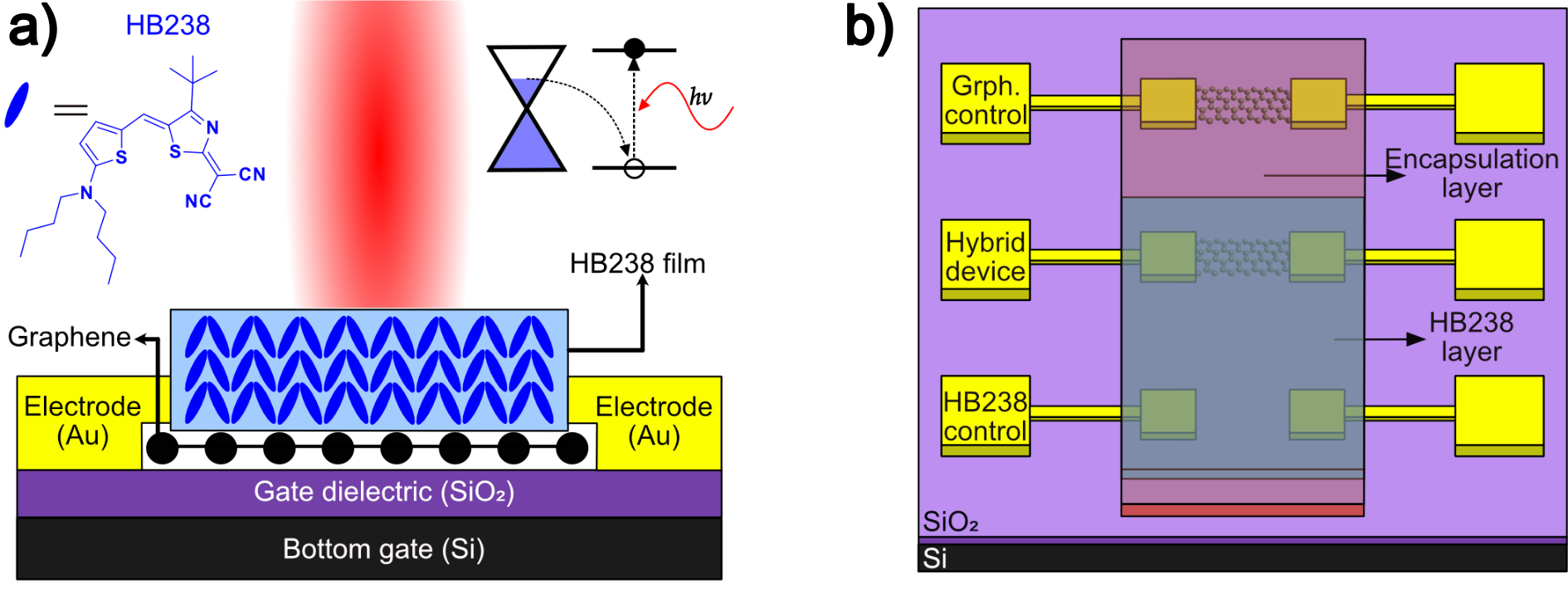}
    \caption{Molecular aggregate enhanced photodetection. a) Light is absorbed in a layer of aggregates of the merocyanine HB238. Photogenerated carriers are injected into the graphene layer acting as the active material in a phototransistor. The photodoping results in photodetection. b) We fabricate on the same chip hybrid devices combining molecular aggregates with graphene as well as controls with only molecular aggregates or graphene. All devices are encapsulated to protect them under ambient conditions.}\label{fig:concept}
\end{figure}

The concept of our study is illustrated in Fig.~\ref{fig:concept}a. Thin films of the merocyanine 2-[5-(5-dibutylamino-thiophen-2-yl-methylene)-4-\textit{tert}-butyl-5\textit{H}-thiazol-2-ylidene]-malononitrile  (from hereon referred to as HB238\cite{buerckstuemmer2011}) act as light harvesting element in a hybrid graphene phototransistor. The optical and charge transport properties of HB238 have been studied in detail by us and others.\cite{Liess2017,liess2019,buerckstuemmer2011,Schembri2021,schaefer2024,boehner2024} Due to strong intermolecular interactions, HB238 readily aggregates in the solid state. In thin films of HB238, the molecules aggregate in an oblique manner as illustrated in Fig.~\ref{fig:concept}a. This results in Davydov splitting with J- and H-like transitions.~\cite{boehner2024} On substrates the aggregate layers display uniaxial anisotropy with the transition dipole moment of the H-transition being oriented out-of-plane and that of the J-transition lying in the sample plane.~\cite{boehner2024} This results in strongly light-absorbing thin films with narrow absorption bands that can be selectively addressed using the polarization of light. We apply this in photodetection by depositing  15~nm of HB238 on bottom-gated graphene transistors (see Fig.~\ref{fig:concept}a). Full details on the sample fabrication are given in the Methods section. The deposition of HB238 takes place by physical vapor deposition on a heated substrate, which results in aggregation of the molecules on the target substrate. 

To demonstrate and study the influence of the layer of molecular aggregates on the photoresponse of the graphene phototransistor, we employ a novel fabrication technique that enables the integration of multiple, identical devices on a single chip. The device layout is shown in Fig.~\ref{fig:concept}b. The novel device is designated `hybrid device' and consists of a graphene transistor sensitized with the HB238 aggregate layer. To study the influence of the aggregates on the photoresponse and charge transport properties of graphene, we prepare control devices with only HB238 (`HB238 control') or graphene (`graphene control'), see Fig.~\ref{fig:concept}b. The graphene devices are identical and the film of HB238 aggregates covering the devices is homonogeneous. Several devices can be prepared on the same chip allowing better statistics and parameter studies to be carried out. Full details on the sample preparation are given in the Methods section and the Supporting Information. 

\begin{figure}
    \centering
    \includegraphics[width=14cm]{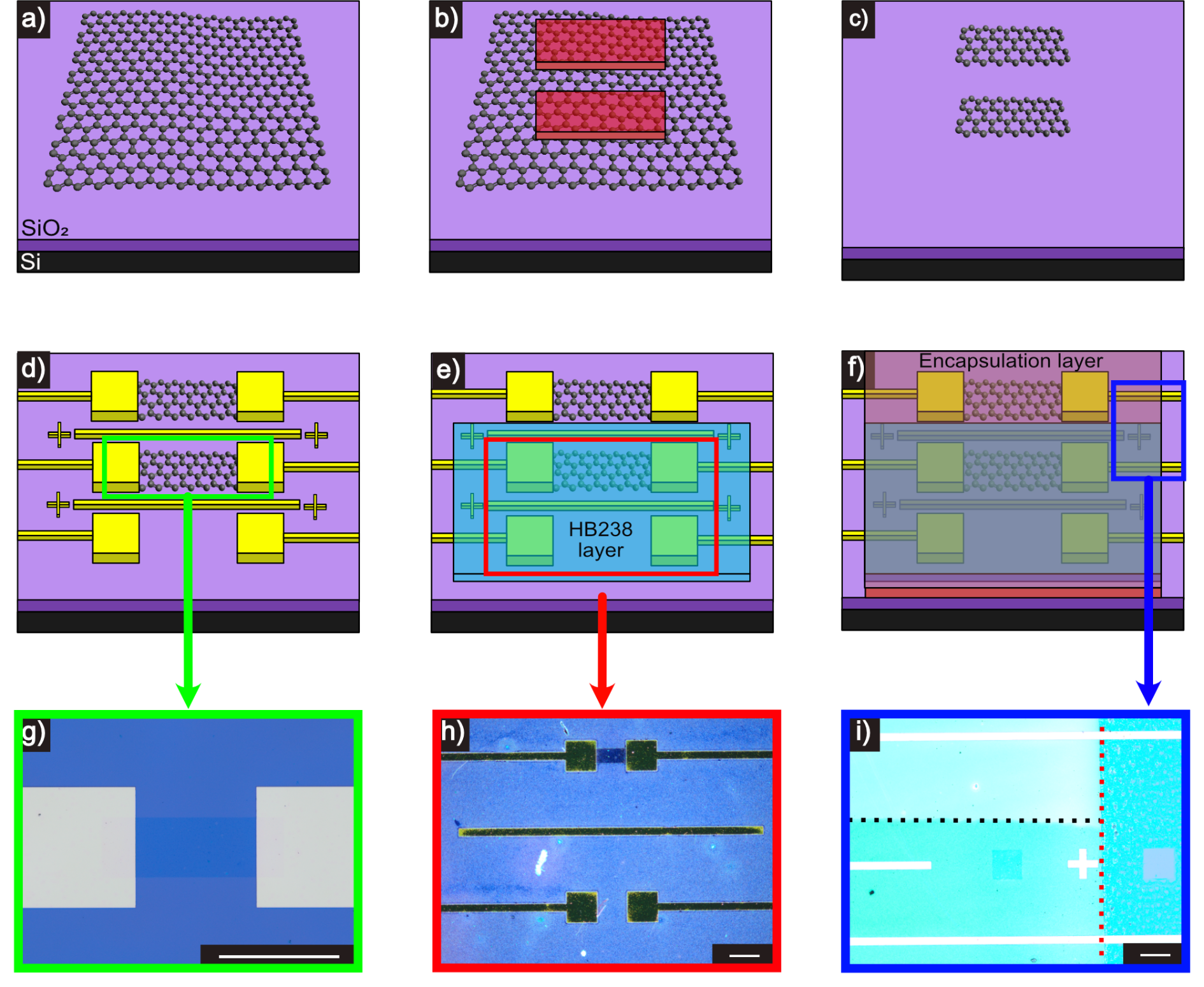}     
    \caption{Sample fabrication. a) CVD grown graphene is transferred onto a transistor substrate. b) Laser lithography is used to define an etch mask c) followed by reactive ion etching to define graphene patches. d) Electrodes are patterned using laser lithography, metal deposition, and a lift-off process. e) Selected devices are covered by a layer of aggregated HB238 molecules by physical vapor deposition. f) The devices are completed by encapsulating the active regions to shield them from ambient conditions. g) Optical micrograph of the green rectangle in (d). h) Optical micrograph of the red rectangle in (e). i) Optical micrograph of the blue rectangle in (f).  All scale bars are 200 $\mu$m.}\label{fig:fabrication}
\end{figure}

The key steps of our sample fabrication are illustrated in Fig.~\ref{fig:fabrication}. In short, CVD grown graphene is wet transferred onto Si/SiO$_2$ substrates. The silicon substrate is highly doped allowing bottom gating of the transistor. The graphene channels for the transistors are patterned using laser lithography and reactive ion etching (Fig.~\ref{fig:fabrication}b,c). Subsequently gold electrodes and contact pads are fabricated onto the substrate using laser lithography and thermal evaporation followed by a lift-off process (Fig.~\ref{fig:fabrication}d). The overcoating of the etched edges of graphene with metal most likely results in edge contacts between the electrodes and graphene, which offer superior charge transport properties.~\cite{Huang2023} To coat only selected devices with HB238 we mask the sample using a shadow mask leaving exposed only those graphene areas, which are to be sensitized by the molecular aggregates (Fig.~\ref{fig:fabrication}e). Similarly, the contact pads are masked to avoid deposition of HB238 on them. To provide a clean interface between graphene and HB238, the device is kept under high-vacuum overnight before deposition of the organic layer to desorb any absorbed water or oxygen. Following the HB238 evaporation, the sample is encapsulated in a trilayer stack of Parylene C/Al$_2$O$_3$/Parylene C (see Methods for full details). Finally, the mask is removed from the contact pads to allow unobstructed contacting of the probe needles for optoelectronic measurements. 

Figure~\ref{fig:fabrication}g is an optical micrograph of the green box illustrated in Fig.~\ref{fig:fabrication}d, showing a bright-field image of a graphene channel and gold electrodes. Figure~\ref{fig:fabrication}h is a dark-field optical micrograph of transistor devices with and without graphene as illustrated by the red box in Fig.~\ref{fig:fabrication}e. Figure~\ref{fig:fabrication}i is a dark field optical micrograph of the blue box illustrated in Fig.~\ref{fig:fabrication}f which shows the intersection of the regions with and without the dye sensitization (separated horizontally in Fig.~\ref{fig:fabrication}i by the black dotted line) as well as the regions with and without the encapsulation layer (separated vertically by the red dotted line in Fig.~\ref{fig:fabrication}i). 

It is noteworthy that after keeping the sample overnight under high vacuum before depositing HB238, the devices are not exposed to ambient conditions until removal from the Parylence C deposition chamber. As a consequence of this, the substrate remains protected from ambient conditions by virtue of the encapsulation layer. The encapsulation is essential for avoiding doping of graphene by impurities. Surprisingly, we find that devices covered only by HB238 or encapsulated by thin layers of Parylene C display ambient degradation (data not shown here). This illustrates the extreme sensitivity of graphene to external disturbances and the need for proper shielding.

\subsection*{Photodoping and transfer characteristics}

\begin{figure}
    \centering
    \includegraphics[width=16cm]{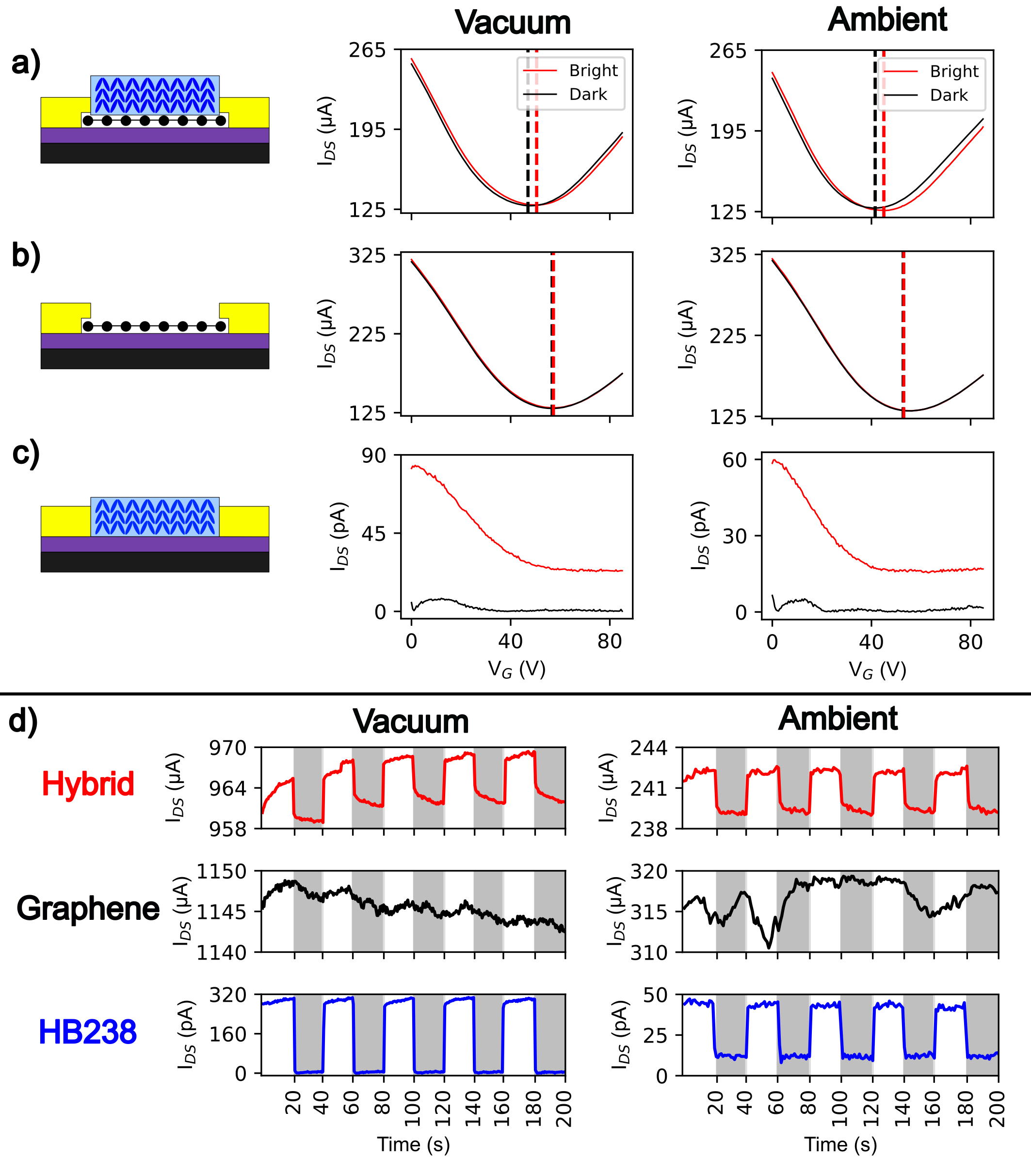}
    \caption{Transfer characteristics. The transfer curves for a) hybrid device, b) graphene control, and c) HB238 control are shown for measurements under vacuum and ambient conditions. d) The drain current modulation for the devices under vacuum and ambient conditions are displayed. 
    }\label{fig:transfercurves}
\end{figure}

We now turn to the electrical characterization of the devices. The devices are characterized both under vacuum and ambient conditions to test the the encapsulation procedure. The transfer curves for the three types of devices under vacuum  and ambient conditions, with and without illumination are shown in Fig.~\ref{fig:transfercurves}a-c. The devices with graphene channels (hybrid device and graphene control, Fig.~\ref{fig:transfercurves}a and b, respectively) show typical transfer characteristics with Dirac points close to 50~V. This p-doped nature of the graphene channels is expected due to residual dopants from the various processing steps involved in the fabrication process. 

The data in Fig.~\ref{fig:transfercurves}c corresponds to the device consisting just of HB238 between the source and drain electrodes and displays the typical transfer curve for an HB238 organic field effect transistor (OFET) with low drain current.  Here, it is noteworthy that the transfer characteristics of all three device types under vacuum are almost identical to their counterparts measured under ambient conditions, thereby indicating successful encapsulation of the devices to protect from passive doping caused by exposure to ambient conditions. Degradation of the transfer curves in the absence of encapsulation due to ambient exposure is illustrated in Supporting Information section S1 and is comparable to that observed for other studies on graphene transistors.\cite{Rasouli2023}

Upon illumination, a rightward shift of the transfer curve is observed for the hybrid device indicating a further p-doping of the graphene channel upon photoexcitation of the HB238 layer (Fig.~\ref{fig:transfercurves}a). The optical density of an HB238 film of the thickness used here is approximately 0.075 at 700~nm wavelength, thereby implying a significant enhancement in light absorption compared to monolayer graphene. This p-doping implies the formation of an interfacial electric field due to the transfer of photogenerated charge carriers from the HB238 layer. This photodoping is absent for the pristine graphene control, which is expected due to the poor light absorption of graphene. This confirms that the photodoping observed in the hybrid device is due to light absorption in the HB238 layer. For the HB238 control, the comparison between the bright and dark states under vacuum shows a drain current modulation of approximately 0.1~nA at a gate voltage of $V_\mathrm{G}=0$~V. In contrast, the photodoping induced by the light absorption in the HB238 layer results in a drain current modulation of approximately 10~$\mu$A in the hybrid device (also at $V_\mathrm{G}=0$~V). This observation coupled with the fact that the photodoping effect is observed only in the presence of HB238 together acts as a proof-of-concept for the operating principle of our hybrid device. Due to the small current magnitudes in the HB238 control device, the standard electrode design (shown here in Fig.~\ref{fig:fabrication}) is replaced with an interdigitated electrode pattern for improved signal to noise ratio (details in Supporting Information section S6). Further statistics about the transfer characteristics of the hybrid devices and the graphene controls are given in Supporting Information section S2.

To further study the photoinduced drain current modulation, we cycle the illumination to observe the resulting current modulation. The data is shown in Fig.~\ref{fig:transfercurves}d (illumination is cycled in 20 second intervals with grey portions showing dark state and white portions showing bright state). Upon illumination, a photocurrent is observed for the hybrid and HB238 control devices. However, the baseline current in the graphene channel is significantly higher due to the better carrier mobility of graphene compared to HB238. For the graphene control device, no observable photoresponse is visible, as is expected because of the poor light absorption of monolayer graphene. The hybrid device's photocurrent modulation of 10~$\mu$A is enormously enhanced from that of the graphene control device where no clear signal is visible. Compared to the response of the HB238 control, we conclude that the hybrid device results in a 10$^{5}$ enhancement of the signal. These results demonstrate the motivation behind our hybrid device where the high light absorption of HB238 is combined with the electrical performance of graphene.

Lastly, to further test the quality of our encapsulation protocol, the illumination cycling experiments conducted in vacuum were repeated under ambient conditions. Similar to the transfer characteristics, these experiments also show identical responses to their counterparts measured in vacuum. The encapsulation of the devices thus provides sufficient shielding and allows further studies outside inert environments. The absolute magnitudes of the drain current ($I_\mathrm{DS}$) for the ambient conditions is lower than its vacuum counterparts as a lower drain source voltage ($V_\mathrm{DS}$=2~V) was used under ambient conditions to minimize resistive heating effects, as the objective was purely to show the functioning of the encapsulation procedure. Under vacuum a higher drain source bias ($V_\mathrm{DS}$=5~V) was used for improved signal-to-noise. 

\subsection*{Origin of photodoping}

\begin{figure}
    \centering
    \includegraphics[width=16cm]{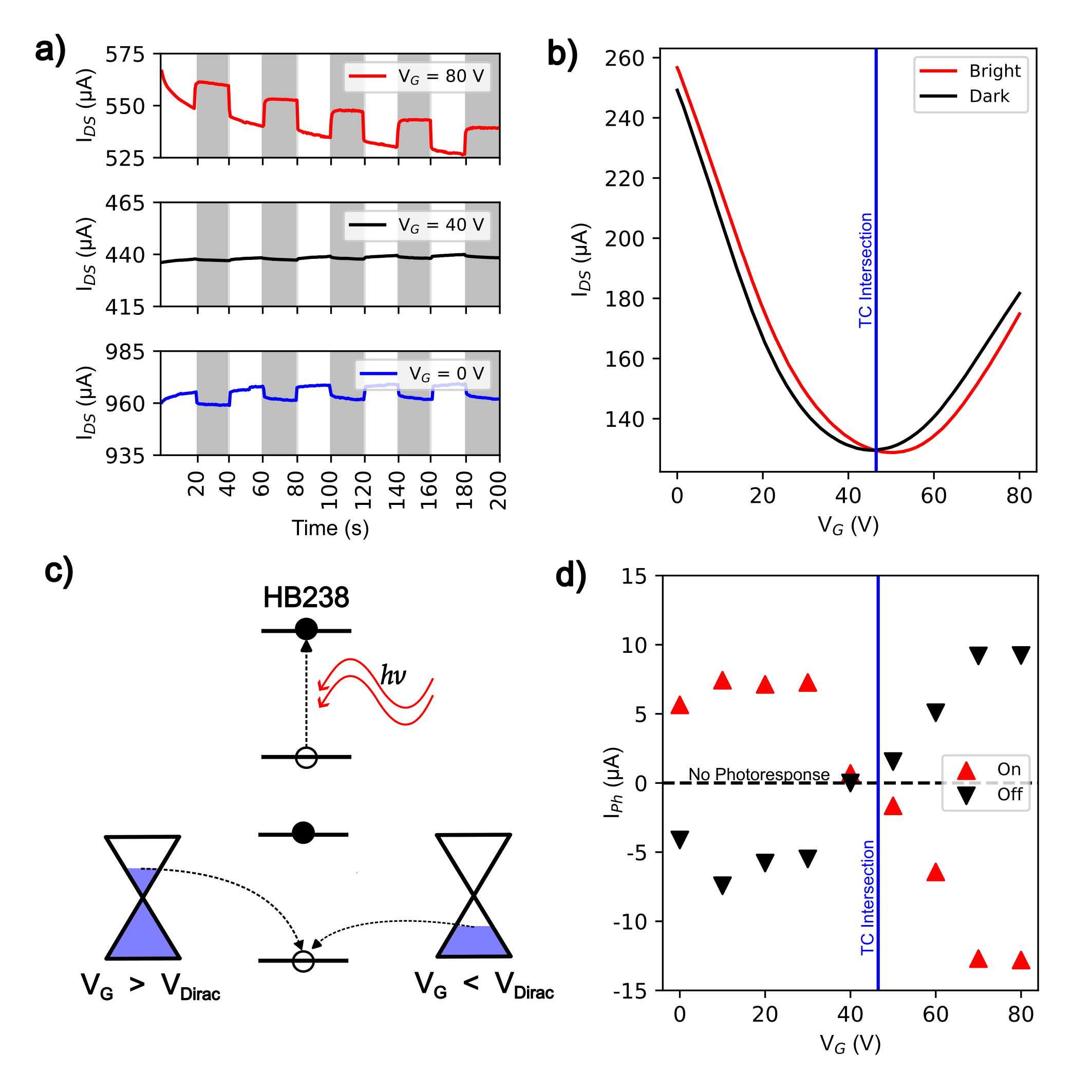}
    \caption{Influence of gate voltage on photocurrent. a) Photocurrent modulation for three different values of the gate voltage. b) Transfer curve of the hybrid device considered in panel a. c) Schematic of the photodoping process. d) Photocurrent as a function of gate voltage. The values labeled `on' correspond to the amplitude of the photocurrent when the light is switched on while the values labeled `off' refer to the change in current when the illumination is turned off. The blue line in b and d marks the gate voltage at which the transfer curves for bright and dark states intersect.}\label{fig:cycling}
\end{figure}

To further investigate the dynamics of photogenerated charge carriers in the hybrid device, we next measure the drain current cycling as a function of gate voltage $V_\mathrm{G}$. Figure~\ref{fig:cycling}a shows the on/off cycles for a device (transfer curve in Fig.~\ref{fig:cycling}b) at three different gate voltages. The Dirac points ($V_\mathrm{Dirac}$) of most hybrid devices are close to 40~V. This value for the gating thus brings the device close to charge neutrality. As can be seen in Fig.~\ref{fig:cycling}a, at $V_\mathrm{G}=80$~V ($V_\mathrm{G}>V_\mathrm{Dirac}$) illumination of the sample results in an decrease in the drain current ($I_\mathrm{DS}$), i.e., a negative photocurrent ($I_\mathrm{Ph}$<0). In this gate voltage regime, the dominant charge carriers in the graphene channel are electrons. A negative photocurrent upon illumination indicates a decrease in the number of free charge carriers in the graphene channel, thereby implying that photogenerated holes from the HB238 layer are injected into the graphene channel.

At $V_\mathrm{G}=40$~V ($V_\mathrm{G}=V_\mathrm{Dirac}$), the magnitude of the generated photocurrent is minimized. This is in agreement with the response observed in the transfer curves for the hybrid device at 40~V where the drain current modulation is minimized (see Fig.~\ref{fig:cycling}b). On the other hand, when $V_\mathrm{G}=0$~V ($V_\mathrm{G}<V_\mathrm{Dirac}$) illumination of the sample results in a positive photocurrent. This implies that even with a reduction in the Fermi level of the graphene channel, the mechanism of photogenerated charge transport remains unchanged, i.e., photogenerated holes are transferred from the HB238 layer to graphene irrespective of the position of the Fermi level of the graphene. Furthermore, it is noticeable that the magnitudes of the generated photocurrent is higher at $V_\mathrm{G}=80$~V ($V_\mathrm{G}>V_\mathrm{Dirac}$) than at  $V_\mathrm{G}=0$~V ($V_\mathrm{G}<V_\mathrm{Dirac}$). This effect can be attributed to the fact that as the Fermi level of the graphene channel rises with an increase in gate voltage, the transfer of an electron (from graphene) to satisfy the vacancy generated in the HB238 layer upon photoexcitation becomes more energetically favorable, as schematically illustrated in Fig.~\ref{fig:cycling}c. To elucidate this phenomenon, we measure the influence of the gate voltage $V_\mathrm{G}$ on the photocurrent $I_\mathrm{Ph}$ as shown in Fig.~\ref{fig:cycling}d. Here, it is evident that the amplitudes of the generated photocurrent is highly dependent on the gate voltage applied. The generated photocurrent changes sign at approximately 46~V which is in proximity to the voltage at which the transfer curves for the bright and dark states of the hybrid device intersect (as shown by the blue line in Fig.~\ref{fig:cycling}b). This facet of being able to adjust not only the magnitude but also the sign of the generated photocurrent affords an additional facet of multimodal tunability to the hybrid device. Supporting Information section S3 shows further statistics on the illumination cycling experiments for the hybrid devices as well as graphene controls, before and after HB238 deposition.

\subsection*{Spectral photoresponse}

\begin{figure}
    \centering
    \includegraphics[width=16cm]{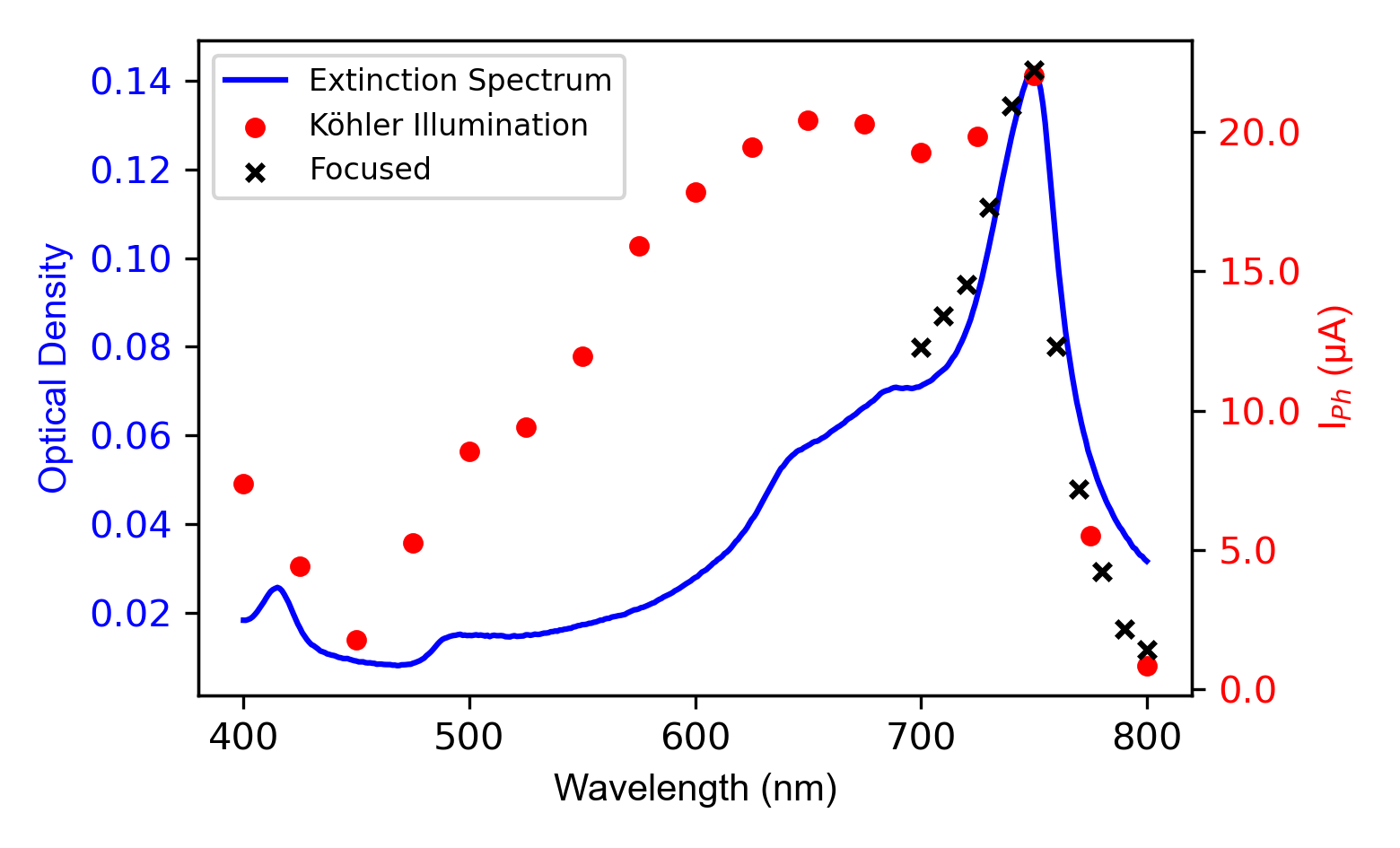}
    \caption{Spectral photoresponse. The photocurrent (circles and crosses) has a similar wavelength dependence as the extinction (blue line). The signal peaks around 750 nm at the J-transition.}\label{fig:spectroscopy}
\end{figure}

We next turn to the spectral properties of the photoresponse. For this we illuminate the hybrid device with light from a tunable laser. The incident light is only weakly focused (K\"ohler illumination) so that we obtain the average photoresponse over a large area. Details of the optical characterization are given in the Methods section. Figure~\ref{fig:spectroscopy} shows the spectral photocurrent together with the extinction spectrum of HB238. The extinction spectrum was measured from a thin film prepared on a glass substrate in the same deposition run as the phototransistor. Supporting Information section S5 shows angle-resolved spectra of the film on glass substrate. In general, the photocurrent generated by the weakly focused light is well correlated to the extinction spectrum. The response peaks at the extinction maximum at approximately 750~nm which corresponds to the J-transition. We remark that the H-transition at 490~nm is not visible here due to the normally incident light, which cannot excite this transition due to the normally oriented transition dipole moment.~\cite{boehner2024,schaefer2024} However, it is also noticeable that the photoresponse in the spectral range of 550--700~nm relative to the extinction is significantly higher than in the 700--800 nm range. The shoulder on the shorter wavelength side of the J-aggregate peak is due to non-aggregated monomers or small aggregates (see Supplementary Fig.~S5). It is possible that the transfer of photogenerated charge carriers to graphene is more likely from such monomers or small aggregates than from larger aggregates which support more delocalized states. Figure ~\ref{fig:spectroscopy} also shows the spectrally resolved photoresponse at higher spectral resolution around the J-transition for focused illumination. Again, the generated photoresponse is well correlated to the extinction spectrum. However, when comparing this to the weakly focused illumination, a distinct difference is observed. At the wavelengths in the range of 700 - 750~nm the photoresponse for focused light aligns more closely to the extinction spectrum than its weakly focused counterpart. The correlation between generated photoresponse and the extinction spectrum appears to depend on the focusing conditions. This effect is noteworthy and warrants further investigation. 

\begin{figure}
    \centering
    \includegraphics[width=16cm]{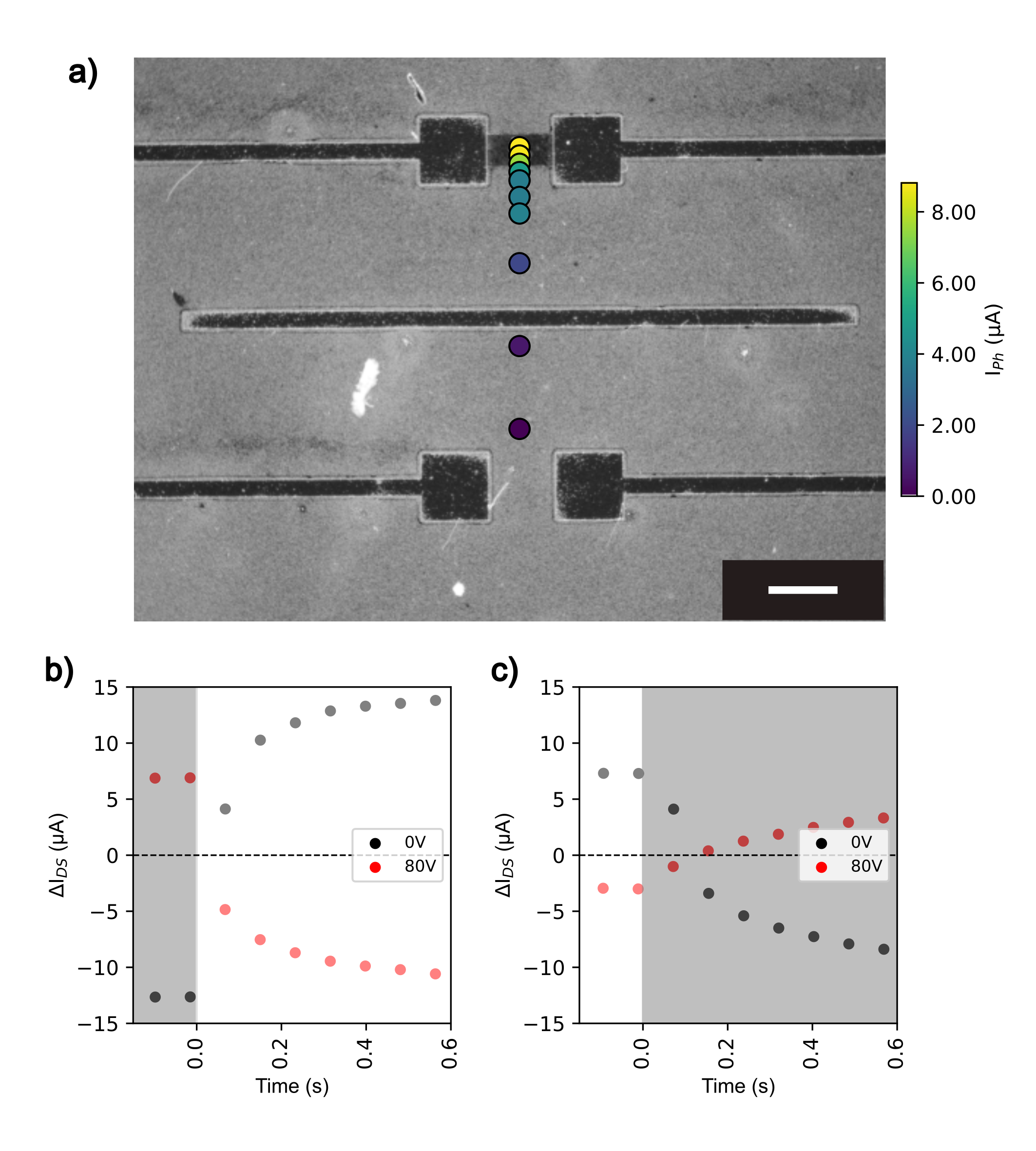}
    \caption{Spatial and temporal photoresponse. a) The strength of the photocurrent as a function of position on the sample. The microscope image shows the position of the measured photocurrent. The scale bar is 200~$\mu$m. Photoresponse when the light is turned on b) and off c) for gate voltages below (0~V) and above (80~V) the Dirac point.}\label{fig:spatial}
\end{figure}

To gain more insight into the photogenerated charge carrier transport, we spatially resolve the generated photoresponse by illuminating the sample locally with focused light. For optimal signal-to-noise ratio, we excite the device at the photocurrent (extinction) maximum of 750~nm. Figure~\ref{fig:spatial}a shows the photocurrent as a function of position on the sample. Due to the nature of the device layout, the HB238 film extends beyond the graphene channel and therefore photocurrent generation can be observed at distances up to 800~$\mu$m from the channel edge. The generated photocurrent is maximised within the graphene channel and decays as the distance from the channel edge increases. The long range of the photoresponse is surprising and indicates that the photodoping can take place nonlocally. Our data clearly demonstrates that neither the graphene nor the HB238 control devices results in significant photocurrent. The observed long range response thus originates from the combination of the two. This interaction should be studied in more detail in the future.

Lastly, Fig.~\ref{fig:spatial}b-c show the temporal response of the device. Here we distinguish between switching on and off situations and show the data for gating below and above the Dirac point. It is evident that switching on transition is significantly faster than the reverse. This is in agreement with the data in Fig.~\ref{fig:cycling}d in which the `on' steps are clearly larger in magnitude than the `off' steps. This has been observed in previous studies on hybrid organic phototransistors and has been attributed to the trapping of photogenerated electrons in defects or disorders in the organic thin films.\cite{Liu2016,Dai2024,Dong2012} The trapping of such photogenerated electrons can significantly prolong the annihilation times. Studying the temporal response as a function of order and structure of strongly aggregating materials such as HB238 is critical for improved understanding of charge transport in hybrid devices.

\section*{Conclusions}

In summary, we have designed and fabricated a graphene-HB238 hybrid phototransistor. Our concept combines the favorable electrical properties of graphene with the strong and tunable light absorption of HB238, resulting in a hybrid transistor with 10$^{5}$ enhancement of the photoresponse compared to the HB238 control device. The dynamics of photogenerated charge carriers were studied to show tunable photodoping by virtue of adjusting the Fermi level of the graphene channel in the hybrid device, affording control over the sign and magnitude of the generated photocurrent. We study the photoresponse generated in the hybrid device as a function of excitation wavelength. While the spectrally resolved photoresponse is well correlated with the extinction spectrum of the HB238 layer, the results also indicate a dependence on the mode of illumination utilized. We believe this observation provides a pathway for further investigations into the optoelectronic responses of such aggregating organic dyes. A novel fabrication strategy allows for the collection of data from multiple device types on the same chip, to minimize artifacts generated from various device faults or fabrication procedures. Furthermore, our encapsulation protocol enables stable measurement of devices under ambient conditions to afford comprehensive optoelectronic characterization. 

The absorption characteristics of aggregating organic dyes is dependent on a range of parameters such as wavelength, angle of incidence, polarization and state of aggregation, thereby supporting the fabrication of a multimodally tunable hybrid transistor. The multimodally tunable hybrid phototransistor acts as an ideally suited testbed to investigate the photogenerated charge carrier dynamics at the interface between aggregating organic dyes and 2D-templating materials like graphene. Such studies are integral in understanding the charge transport properties of aggregating organic dyes and can be a stepping stone towards significant improvements in device-relevant organic optoelectronic technologies. 

\section*{Methods}

\subsection*{Sample fabrication}

Highly doped silicon wafers with a 290~nm gate oxide were cut into 15~mm~$\times$~15~mm substrates. The substrates were cleaned by ultrasonication at 40~$^{\circ}$C in deionized (DI) water, Mucasol (2 \% aqueous solution), DI water, acetone, and isopropyl alcohol, and then dried by blowing with dry nitrogen. To allow for lithographic alignment across multiple fabrication steps, first markers were patterned onto the substrate using standard laser lithography followed by metal deposition by vacuum evaporation and a lift-off process. After the marker patterning, graphene was transferred by PMMA-assisted wet transfer. For this, CVD grown graphene films on a sacrificial polymer layer (Graphenea EasyTransfer) were cut to appropriate sizes and delaminated into ultrapure water. The graphene/PMMA layer was then fished out onto the substrate and allowed to dry at a 45$^{\circ}$ angle at room temperature for 1 hour, following which the substrate was placed on a hotplate at 140 $^{\circ}$C for 2 hours. The dried graphene layer was stored in a low-moisture environment overnight for better adhesion to the substrate, following which the PMMA layer was washed away in 50~$^{\circ}$C acetone for 2 hours and soaking in isopropyl alcohol at room temperature for 1 hour and then drying with dry nitrogen. The graphene flake was then patterned into the desired channels using laser lithography and reactive ion etching with oxygen plasma. Finally, electrodes and contact pads were  patterned onto the substrate using the same protocol as used for the alignment markers. As metal we used 2~nm of titanium followed by 45~nm of gold. 

\subsection*{HB238 deposition}

The HB238 was evaporated onto the substrate using a rate of 0.03~\AA/s and a substrate temperature of 74~$^\circ$C under high vacuum conditions (pressure approximately $10^{-7}$~mbar).

\subsection*{Encapsulation}

For the encapsulation procedure, 100~nm Parylene C, 50~nm Al$_{2}$O$_{3}$, and 100~nm of Parylene C were sequentially deposited onto the sample. The Parylene C coating was performed by vaporizing the powdered precursor at 130-140~$^{\circ}$C in a parylene coater. The gaseous dimer was then pyrolysed into a monomer at 690~$^{\circ}$C. The polymeric films of Parylene C then formed on the samples in the main chamber of the coating system, which was kept at room temperature and at a base pressure below 25~mTorr. Al$_{2}$O$_{3}$ was deposited by atomic layer deposition (ALD). The trimethylaluminum and H$_{2}$O cylinders connected to the ALD reactor were kept at room temperature. A 50~nm thick Al$_{2}$O$_{3}$ layer was then deposited by 455 cycles of  15~ms trimethylaluminum pulse/10~s N$_{2}$ purge/15~ms H$_{2}$O pulse/10~s N$_{2}$ purge. The ALD was carried out at an operational temperature of 80~$^{\circ}$C and a base pressure of 0.1~Torr. From loading the sample into the vacuum evaporator for the deposition of HB238 until encapsulation, the sample was kept under vacuum or dry nitrogen atmosphere. In order to guarantee good electrical contact, the contact pads were masked with adhesive tape during the entirety of the deposition and encapsulation process. Following encapsulation and removal from the nitrogen atmosphere, the adhesive tape is removed from the contact pads. 

\subsection*{Optoelectronic characterization}

The spectrally broadband electrical characterization of the transistors was performed using a vacuum probe station and a semiconductor parameter analyzer. Measurements in vacuum were done after leaving the sample in vacuum (pressure 10$^{-6}$~mbar) chamber for at least 8 hours to allow for desorption of surface dopants. A halogen lamp was used as the illumination source for measuring the photoresponse. To avoid artifacts produced by possible leakage currents, any graphene channels with gate currents ($I_{G}$) such that $I_{DS}/I_{G} < 10^{2}$ were ignored. Typically for most graphene channels a value of $I_{DS}/I_{G}$ of $10^{3}$-$10^{5}$ was observed.

Spectrally-resolved photoresponses were measured using a femtosecond tunable pulsed laser. Here, the electrical readout from the devices was measured using two source measurement units. One device was used to gate the transistor and measure the gate current. The other unit was used for the source-drain bias and measurement of the drain current. We used focused light obtained using a low numerical aperture microscope objective to collect the position-resolved photoresponse by translating the sample using a precision stage. The intensity of the focused illumination is approximately 1~kW/cm$^2$. For the non-position resolved data we used K\"ohler illumination with an approximately 300~$\mu$m diameter spot with an intensity of approximately 1~W/cm$^2$.

\section*{Funding sources}

This project is funded with support from the RTG-2591 "TIDE - Template-designed Organic Electronics" (Deutsche Forschungsgemeinschaft). K.L. and A.R. acknowledge funding from the DFG project 426882575. Instrument funding by the Deutsche Forschungsgemeinschaft in cooperation with the Ministerium für Kunst und Wissenschaft of North Rhine-Westphalia (projects 448775637 and INST 216/1121-1 FUGG) is acknowledged. M.C.G. thanks the Alexander von Humboldt Foundation for the Humboldt Professorship.
	  
\section*{Acknowledgements}

Research was supported by the University of Cologne through the Institutional Strategy of the University of Cologne within the German Excellence Initiative (QM$^2$).\\

\noindent The authors declare no competing financial interest.

\section*{Author contributions}

A.R. fabricated the samples, characterized the devices, and analyzed the data. M.S. participated in the initial steps of the experimental study. C.-K.M. and M.C.G. contributed to the nanofabrication and M.N. and K.M to the fabrication of the aggregate layers. K.L. planned and supervised the project. All authors contributed to the analysis of the results and writing of the paper.

\begin{suppinfo}

Influence of pressure on transfer characteristics, additional data on photogating and illumination cycling, Raman spectra of graphene on the devices, angle-resolved extinction spectra of HB238, and electrode layout pattern.

\end{suppinfo}

\newpage

%\bibliography{literature.bib}
\providecommand{\latin}[1]{#1}
\makeatletter
\providecommand{\doi}
  {\begingroup\let\do\@makeother\dospecials
  \catcode`\{=1 \catcode`\}=2 \doi@aux}
\providecommand{\doi@aux}[1]{\endgroup\texttt{#1}}
\makeatother
\providecommand*\mcitethebibliography{\thebibliography}
\csname @ifundefined\endcsname{endmcitethebibliography}
  {\let\endmcitethebibliography\endthebibliography}{}

\end{document}